\documentstyle[11pt]{article}

\begin{document}
%

\begin{center}
{\large \bf Neutrino masses and the extra $Z^{\prime 0}$}

\vskip.5cm

W-Y. Pauchy Hwang\footnote{Correspondence Author;
 Email: wyhwang@phys.ntu.edu.tw; arXiv:0808.2091 (hep-ph, August 18, 2008)} \\
{\em Asia Pacific Organization for Cosmology and Particle Astrophysics, \\
Institute of Astrophysics, Center for Theoretical Sciences,\\
and Department of Physics, National Taiwan University,
     Taipei 106, Taiwan}
\vskip.2cm


{\small(September 9, 2010; Revised May 23, 2011)}
\end{center}

\begin{abstract}
We now know firmly that neutrinos have tiny masses, but in the minimal Standard
Model there is no natural sources for such tiny masses. On the other hand, the
extra heavy $Z^{\prime 0}$ requires the extra Higgs field, the particle generating
the $Z^{\prime 0}$ mass and also the neutrino tiny masses. Here I propose that the
previous 2+2 Higgs scenario could be a natural framework for mass generations
for both the neutrinos and the extra $Z^{\prime )}$. These particles are
classified as the "dark matter". Hopefully all the couplings to the
"visible" matter are through the neutrinos and the extra $Z^{\prime 0}$ vis
the so-called "minimum Higgs hypothesis",
explaining naturally why the dark matter is so dark to see.

\bigskip

{\parindent=0pt PACS Indices: 12.60.-i (Models beyond the standard
model); 12.10.-g (Unified field theories and models); 14.70.Pw
(Other gauge bosons).}
\end{abstract}

\section{Introduction}

Neutrinos have masses, the tiny masses. It may be difficult to explain,
since neutrinos in the minimal Standard Model are massless. In fact, this
may be a signature that there is a heavy extra $Z^{\prime 0}$. This extra
$Z^{\prime 0}$ then requires the new Higgs doublet\cite{Hwang}. This Higgs
doublet also generates the tiny neutrino masses.

To think of it more, we may invoke the so-called "minimum Higgs hypothesis"
(to account for the fact that we are yet to see the Higgs particle after
searching for it over the last forty years). So, the extra $Z^{\prime 0}$ and
the required new Higgs doublet would be everything which we should consider, if
there is nothing else. This brings us back to the $2+2$ extra $Z^{\prime 0}$
model\cite{Hwang}. To be honest, it might be too much to assume that "there is
nothing else" - how about the missing the right-handed sector or the family
"gauge" symmetry (to explain way the occurrence of the three generations)?
In this brief report, our minds are so over-simplified that the extra
$Z^{\prime 0}$ plus one new Higgs doublet is the story.

There might be some difficulties as regarding that, in the present Universe, there
is 25\% dark while only 5\% visible ordinary matter - the picture of the extra
$Z^{\prime 0}$ plus the new Higgs doublet may sound too simple and too little
to account for the dark matter world. But we should analyze its consequences since
the situation calls for.

So, the natural clue appears in the message that neutrinos have tiny masses.
The Higgs and extra $Z^{\prime 0}$, so weak and so heavy, would come much later
in the clue (message). They all may belong to the so-called "dark matter", 25\%
of the present-day Universe (compared to 5\% of ordinary matter).

\section{Tiny neutrino masses and the new Higgs doublet}

In a world of point-like Dirac particles interacting with the generalized
standard-model interactions, there are left-handed neutrinos belong to
$SU_L(2)$ doublets while the right-handed neutrinos are singlets.
The term specified by
\begin{equation}
\lambda'\cdot ({\bar\nu}_L,{\bar e}^-) \nu_R \varphi
\end{equation}
with $\varphi=(\varphi^0,\varphi^-)$ the new Higgs doublet could generate
the tiny mass for the neutrino. Let call it the "remote" Higgs doublet.

In the real world, neutrino masses are tiny
with the heaviest in the order of $0.1\, eV$. The electron, the lightest
Dirac particle except neutrinos, is $0.511\, MeV$\cite{PDG} or $5.11 \times 10^5\, eV$.
That is why the standard-model Higgs, which "explains" the masses of all other
Dirac particles, is likely not responsible for the tiny masses of the neutrinos.

In a previous paper in 1987\cite{Hwang}, we studied the extra $Z^{\prime 0}$
extension paying specific attention to the Higgs sector - since in the minimal
Standard Model the standard Higgs doublet $\Phi$ has been used up by $(W^\pm,\,Z^0)$.
We worked out by adding one Higgs singlet (in the so-called 2+1 Higgs scenario) or
adding a Higgs doublet (the 2+2 Higgs scenario). It is the latter that we could add the
neutrino mass term in naturally, according to "the minimum Higgs hypothesis".
(See Ref.\cite{Hwang} for details. Note that the
complex conjugate of the second Higgs doublet there is the $\varphi$ above.)

The new Higgs potential follows the standard Higgs potential, except
that the parameters are chosen such that the masses of the new Higgs are much
bigger. The coupling between the two Higgs doublets should not be too big to
upset the nice fitting\cite{PDG} of the data to the Standard Model. All
these go with the smallness of the neutrino masses. Note that spontaneous symmetry
breaking happens such that the three components of the standard Higgs get absorbed
as the longitudinal components of $W^\pm$ and $Z^0$.

To be specific, let the vacuum expectation values be $v$ and $v'$, as determined by
the Higgs potentials. The couplings of the standard Higgs to quarks and charged
leptons are of order $O(\lambda)$. The couplings of neutrinos to the "remote" Higgs
are of order $O(\lambda')$, which would be $(v/v')^2\cdot O(\lambda)$ in view of the
"minimum Higgs hypothesis".

In other words, the details about the mass generation, through the couplings of
Higgs to the various particles, should be the next important question in the
Standard Model. The "minimum Higgs hypothesis" offers us an avenue to simplify
our thinking in this direction. First, we think that the reason for the tiny masses
of neutrinos is an important clue - it does not come from the standard Higgs since
otherwise the range of what would be given by the standard Higgs is of order
$10^{15}$, too wide the range. Second, for the quark sector, the standard Higgs 
should be adequate according to the minimum requirement - see the next paragraph 
for more. Thirdly, neutrinos should not couple to the standard Higgs, but yes to 
the remote Higgs. Strangely enough, all these come from the "minimum Higgs 
hypothesis".

We could also say something about the cancelation of the
flavor-changing scalar neutral quark currents. Suppose that we work with two
generations of quarks, and it
is trivial to generalize to the physical case of three. We would write
\begin{eqnarray}
({\bar u}_L,\,{\bar d}^\prime_L)d^\prime_R\Phi + c.c.;\nonumber\\
({\bar c}_L,\,{\bar s}^\prime_L)s^\prime_R\Phi + c.c.;\nonumber\\
({\bar u}_L,\,{\bar d}^\prime_L)u_R \Phi^*+c.c.;\nonumber\\
({\bar c}_L,\,{\bar s}^\prime_L)c_R \Phi^*+c.c.,
\end{eqnarray}
noting that we use the rotated down quarks and we also use the complex conjugate
of the standard Higgs doublet. This is a way to ensure that
the GIM mechanism\cite{GIM} is
complete. Without anything to do the opposite, I think that it is reasonable to
continue to assume the GIM mechanism.

The question is that why we have only one Higgs doublet doing the job (the GIM
mechanism). The "minimum Higgs hypothesis" may be the answer.

\section{A World of "Point-like" Dirac Particles}

Coming to think about it, we can rephrasing the Standard Model as a world of
point-like Dirac particles with interactions. Dirac, in his relativistic
construction of Dirac equations, was enormously successful in describing the
electron. Quarks, carrying other intrinsic degrees (color), are described
by Dirac equations and interact with the electron via gauge fields. We also know
muons and tau-ons, the other charged leptons. So, how about neutrinos? Our first guess
is also that neutrinos are point-like Dirac particles of some sort (against Majorana
or other Weyl fields). For some reasons, point-like Dirac particles are implemented
with some properties - that they know the other point-like Dirac particles in our
space-time.

So far, this has been true that it is a world of point-like Dirac particles, or quantized
Dirac fields, with interactions, mediated by gauge fields and modulated slightly by Higgs
fields. This phenomenon, which we call "Dirac similarity principle" in my other paper,
reflects a lot about our four-dimensional space-time. In our physical four-dimensional
space-time, our efforts to define "point-like" may stop here. Particle physics serves
to describe the "point-like" Dirac particles in this space-time.

What is also surprising is the role of "renormalizability". We could construct
quite a few such extensions of the minimal Standard Model - the present
extra $Z^{\prime 0}$, the left-right model\cite{Salam}, and the family gauge
theory\cite{Family}; there are more. Apparently, we should not give up though
the road seems to have been blocked.

\section{A Simple Connection to the Unknowns}

It turns out that a world of point-like Dirac particles as described by quantum
field theory (the mathematical language) is the "smallest" physical world. The
interactions are mediated by gauge fields modulated slightly by Higgs fields.
There may be some new gauge fields, such as the extra $Z^{\prime 0}$
extension\cite{Hwang}, the left-right model\cite{Salam}, or the family gauge
symmetry\cite{Family}, or others, but the first clue is the neutrino masses -
and the next clue will be weaker and subtler, and even weaker. Here, besides
the extra $Z^{\prime 0}$, we name the left-right model and the family gauge
symmetry for the reason that the left-right symmetry is some symmetry missing
(but for no reasons) while the duplication of the generations should be for some
reasons.

There are a bunch of dark matter out there - 25\% of the present Universe
compared to only 5\% ordinary matter. We have the Standard Model for the 5\%
ordinary matter and we should expect to have the theory to cover the
25\% dark matter. When we get more knowledge on the dark
matter, we may have pretty good handles of this Universe.

\section*{Acknowledgments}
This research is supported in part by National Science Council project (NSC
99-2112-M-002-009-MY3).


\begin{thebibliography}{99}

\bibitem{Hwang} W-Y. P. Hwang, Phys. Rev. {\bf D36}, 261 (1987). See
the paper for more references.

\bibitem{PDG} Particle Data Group, "Review of Particle Physics",
J. Phys. G: Nucl. Part. Phys. {\bf 33} (2006) 1; on neutrino mass
and mixing, see pp. 156 - 164; see also {\bf 37}, 164 (2010).

\bibitem{GIM} S.L. Glashow, J. Iliopoulos, and L. Maiani, Phys. Rev.
{\bf D2}, 1285 (1970).

\bibitem{Salam} J.C. Pati and A. Salam, Phys. Rev. {\bf D10}, 275 (1974);
R.N. Mohapatra and J.C. Pati, Phys. Rev. {\bf D11}, 566 (1975); {\bf D11},
2559 (1975).

\bibitem{Family} W-Y. Pauchy Hwang, International J. Mod. Phys.
{\bf A24}, 3366 (2009).

\end{thebibliography}
\end{document}